\documentclass[%
 reprint,
 amsmath,amssymb,
 aps,
 pra]{revtex4-2}
\usepackage{float}
\makeatletter
\let\newfloat\newfloat@ltx
\makeatother
\usepackage{algorithm}
\usepackage{algpseudocode}
\usepackage[hidelinks]{hyperref}
\usepackage{amsmath}
\usepackage{dsfont}
\usepackage{braket}
\usepackage{parskip}
\usepackage{xcolor}
\usepackage{graphicx}% Include figure files
\usepackage{dcolumn}% Align table columns on decimal point
\usepackage{bm}% bold math
\usepackage{silence}
\DeclareMathOperator{\tmsv}{TMSV}
\DeclareMathOperator*{\ooo}{o}
\DeclareMathOperator{\identity}{\mathds{1}}
\NewDocumentCommand\petito{o m}{
  \IfNoValueTF{#1}{
    \ooo\left(#2\right)
  }{
    \ooo_{#1}\left(#2\right)
  }
}
\begin{document}

\preprint{APS/123-QED}
\newcommand{\yu}[1]{\textcolor{orange}{#1}}

\title{Displaced Gaussian Boson Sampling for enhanced max-clique search}

\author{Ewan Mer}\email{ewan.mer20@imperial.ac.uk}
\affiliation{Blackett Laboratory, Department of Physics, Imperial College London, London SW7 2AZ, United Kingdom}
\affiliation{Centre for Quantum Engineering, Science and Technology (QuEST), Imperial College London, London SW7 2AZ, United Kingdom}

\author{Zhenghao Li}
\affiliation{Blackett Laboratory, Department of Physics, Imperial College London, London SW7 2AZ, United Kingdom}
\affiliation{Centre for Quantum Engineering, Science and Technology (QuEST), Imperial College London, London SW7 2AZ, United Kingdom}

\author{Shang Yu}
\affiliation{Blackett Laboratory, Department of Physics, Imperial College London, London SW7 2AZ, United Kingdom}
\affiliation{Centre for Quantum Engineering, Science and Technology (QuEST), Imperial College London, London SW7 2AZ, United Kingdom}

\author{Ian A. Walmsley}
\affiliation{Blackett Laboratory, Department of Physics, Imperial College London, London SW7 2AZ, United Kingdom}
\affiliation{Centre for Quantum Engineering, Science and Technology (QuEST), Imperial College London, London SW7 2AZ, United Kingdom}

\author{Raj B. Patel}\email{raj.patel1@imperial.ac.uk}
\affiliation{Blackett Laboratory, Department of Physics, Imperial College London, London SW7 2AZ, United Kingdom}
\affiliation{Centre for Quantum Engineering, Science and Technology (QuEST), Imperial College London, London SW7 2AZ, United Kingdom}

\date{\today}

\begin{abstract}
 Gaussian Boson Sampling (GBS) is capable of solving certain classes of graph problems owing to the samples produced by such a device having a connection to the hafnian matrix function.
 In particular, a GBS device has been shown to provide an enhancement in the search of cliques---or complete subgraphs---in undirected weighted graphs over classical algorithms.
 A graph can be mapped to a GBS experiment by configuring the squeezing parameters of the input states and programming the linear optical network.
 In practice, limited squeezing and photon loss degrade the performance of the GBS device for max-clique search.
 In comparison, coherent states---often considered a classical resource due to their Poissonian statistics---can be readily prepared across many modes using an attenuated laser.
 In this paper, we report an enhancement of the success rate of GBS in finding maximum weighted cliques by adding displacements under lossy conditions or when a limited amount of squeezing is available.
 Moreover, we report that this enhancement can be scaled up to large graphs with limited resource overheads.
 \end{abstract}

\maketitle

\section{\label{sec:level1}Introduction}

Gaussian Boson Sampling (GBS)~\cite{Hamilton2017GaussianSampling, Kruse2019DetailedSampling} has emerged as a means of demonstrating quantum advantage on photonic hardware~\cite{Zhong2020QuantumPhotons,Zhong2021Phase-ProgrammableLight,Madsen2022QuantumProcessor,Liu2025RobustSampling} where deterministically produced single-mode squeezed vacuum (SMSV) states interfere in a Haar-random multimode interferometer followed by photon number measurements across all output modes. Leveraging spontaneous parametric down-conversion in nonlinear materials alleviates the issue of scalability present in Boson Sampling while maintaining the computational hardness of the sampling task. Moreover, GBS has been linked to several practical applications in graph theory and quantum chemistry~\cite{BagheriNovir2025ApplicationsProblems}.

The main challenge to reach the regime of quantum advantage lies in scaling up to a large number of modes whilst achieving a high level of squeezing with low end-to-end loss to overcome the benchmarks set by tensor network algorithms~\cite{Oh2024ClassicalSampling,Liu2023SimulatingOperators,Liu2025RobustSampling,Cilluffo2024SimulatingPicture}. More specifically, a sufficient condition was derived on the level of loss that can be tolerated in the asymptotic limit before the system can be efficiently simulated classically~\cite{Qi2020RegimesSampling}.

The GBS framework has garnered much interest due to its diverse real-world applications, including the computation of vibronic spectra~\cite{Huh2017VibronicTemperature,Jnane2021AnalogSpectroscopy}, topological data analysis~\cite{Yu2025TopologicalProcessor}, and graph applications by embedding data in the unitary of the interferometer and the squeezing parameters. Graph applications include dense-$k$ subgraph search~\cite{Arrazola2018UsingSubgraphs, Sempere-Llagostera2022ExperimentallyDevice} and max-clique search~\cite{Banchi2020MolecularSampling, Yu2023ADiscovery,Yu2026ExtensibleNonlinearity}.

More recently, a variant of GBS called Displaced GBS (D-GBS) has emerged, where displacement gates are applied to each output mode of the interferometer~\cite{Hamilton2017GaussianSampling,Thekkadath2022ExperimentalDisplacement,Li2025ASampling}. These operations can be implemented by interfering coherent states from a laser with each mode of a multi-mode squeezed state on an unbalanced beamsplitter. The parameters of the displacement operation can be readily adjusted by tuning the phase and amplitude of coherent states using standard optical components. Moreover, it has been conjectured that the computational complexity of D-GBS is similar to the complexity of GBS in the low-displacement regime~\cite{Li2025ASampling}. Adding displacement is a requirement for quantum chemistry applications such as computing Franck-Condon factors for vibronic transitions of molecules~\cite{Huh2015BosonSpectra,Huh2017VibronicTemperature} and graph similarity~\cite{Schuld2020MeasuringSampler}.

Here, we show that displacement can also be used to boost the GBS sampling probability of maximum-weighted cliques of a graph under realistic conditions, such as high loss or when the required squeezing levels are unattainable. We also show that this scheme scales as the number of modes increases, a key requirement for potential real-world applications beyond proof-of-concept. First, we review the main properties of D-GBS. Then, we show how adding displacement to GBS can enhance the performance of the GBS-max-clique algorithm.

\begin{figure}[t!]
    \label{fig:GBS_schematic}
    \centering
    \includegraphics{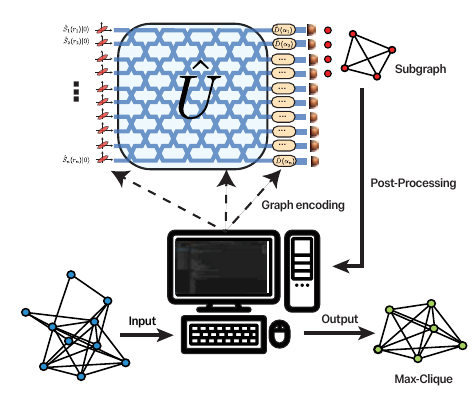}
    \caption{Displaced Gaussian Boson Sampling for max-clique search.
 A weighted graph $G=(V,E)$ is mapped to a set of $(\boldsymbol{r},\boldsymbol{\gamma},\mathcal{U})$ where $\boldsymbol{r}$ and $\boldsymbol{\gamma}$ are the vectors of squeezing parameters and displacements, respectively, and $\mathcal{U}$ represents the passive unitary operation.
 Photons are detected with click-detectors at each output mode. Each photon click pattern is considered a subgraph $S\subseteq G $ and is fed into a greedy shrinking/local search algorithm.
 }

\end{figure}

\section{\label{sec:level2}Theory}
\subsection{\label{sec:level2a}Displaced Gaussian Boson Sampling}
For an $M$-mode GBS experiment with pure states, we denote a measured sample by vector $\boldsymbol n=(n_{1},n_{2},\dots,n_{M})$, where $n_i$ is the number of photons in mode $i$.
 The probability of measuring a specific sample $\boldsymbol{n}$ is related to the hafnian function:
\begin{equation}\label{eq:1}
    p(\bm{n})=\frac{1}{\sqrt{\det(\boldsymbol{\Sigma}_Q)}\,\bm{n}!}|\operatorname{Haf}( B_{\bm{n}})|^{2}
\end{equation}
for $\bm{n}! = n_{1}! n_{2}! \dots n_{M}!$ and $\boldsymbol{\Sigma}_{Q}=\boldsymbol{\Sigma}+\identity/2$, where $\boldsymbol{\Sigma}$ is the covariance matrix of the Gaussian state and $\identity$ is the identity matrix.
 Matrix $B$, defined as $B=\begin{pmatrix}0&\identity\\\identity&0\end{pmatrix}(\identity-\boldsymbol{\Sigma}_{Q}^{-1})$, contains all the information to uniquely define a pure-state GBS experiment, that is, the squeezing parameters $\{r_{i}\}_{i \in [1,M]}$ for each mode $i$ and the passive unitary $\mathcal{U}$ implemented by the interferometer.
 These parameters can be computed from $B$ by applying the Takagi-Autonne decomposition: 
\begin{equation}
\label{Takkagi}
B= \mathcal{U}\;\Big(\bigoplus_{i=1}^{M}\tanh{r_{i}}\Big)\;
 \mathcal{U}^{T}    
\end{equation}

Matrix $B_{\bm{n}}$ is constructed by repeating the $i$-th row and column of $B$, $n_{i}$ times.
 One notes that the dimension of $B$ is necessarily even because single-mode squeezed vacuum states (SMSVs) consist of photon pairs.
 The hafnian of a $2M\times 2M$ symmetric matrix $B$ is defined as follows~\cite{Gupt2019TheSampling,Barvinok2017ApproximatingHafnians}:
\begin{equation}
     \operatorname{Haf}(B)=\sum_{\sigma\in \text{PMP}}\prod_{(i,j)\in \sigma}B_{i,j}
\end{equation}
where $\text{PMP}$ is the set of perfect matchings of $2M$ objects.
 When adding displacement $\boldsymbol d$ to the multi-mode Gaussian state after the interferometer, the probability of measuring a sample $\boldsymbol{n}$ now depends on the loop hafnian function $\operatorname{lHaf}$~\cite{Kruse2019DetailedSampling}:
\begin{equation}\label{eq:D-GBS_prob}
    p(\bm{n})=\frac{\exp(-\frac{1}{2}\boldsymbol d^{\dag}\boldsymbol\Sigma_{Q}^{-1}\boldsymbol d)}{\bm{n}!\sqrt{\det(\boldsymbol{\Sigma}_Q)}}|\operatorname{lHaf}(\text{fd}\{B_{\bm{n}},\boldsymbol\gamma\})|^{2},
\end{equation}
 where $\boldsymbol\gamma=\boldsymbol d^{\dag}\boldsymbol\Sigma_{Q}^{-1}$ is a vector of displacements which replaces the diagonal elements of $ B_{\bm{n}}$, and the loop hafnian is defined as
 \begin{equation}
    \operatorname{lHaf}(B)=\sum_{\mathcal{M}\in \text{SPM}(\boldsymbol{n})}\prod_{(i,j)\in \mathcal{M}}B_{i,j},
\end{equation}
where \emph{SPM} is the set of single-pair matchings of the graph $G(V,E)$ where $V$ and $E$ are the set of vertices and edges, respectively.
 \emph{SPM} is comprised of perfect matchings---a subset of edges that connect each vertex once and only once---and loops on vertices corresponding to the diagonal elements of the adjacency matrix.
 The loop hafnian can be expanded as a sum of hafnians:
\begin{equation}\label{lhafnianexpansion}
\begin{split}
     \operatorname{lHaf}(B_{\bm{n}})=\operatorname{Haf}( B_{\bm{n}})+\sum_{j_{1}=1}^{2N-1}\sum_{j_{2}=j_{1}+1}^{2N} & \gamma_{j_{1}}\gamma_{j_{2}}\operatorname{Haf}(B_{\bm{n}-\{j_{1},j_{2}\}})\\
     & +\dots+\prod_{j=1}^{2N}\gamma_{j},
\end{split} 
\end{equation}
where $B_{\bm{n}-\{j_{1},j_{2}\}}$ is obtained by removing rows and columns $j_{1}$ and $j_{2}$ from $B_{\bm{n}}$.
 An adjacency matrix $A$ can be embedded in the $B$ matrix, which fully characterizes a GBS experiment by a rescaling parameter $c$:
\begin{equation}
    B= c\,A
\end{equation}
where $0<c<1/s_{\text{max}}$ and $s_{\text{max}}$ is the maximum singular value of $A$.
 These conditions are necessary to ensure that the eigenvalues computed by the Takagi-Autonne decomposition are bounded between zero and one, such that they can be mapped to physical squeezing parameters.
 $A$ is a symmetric matrix representation of $G(V,E)$. The weight of an edge connecting two vertices, $v_i, v_j \in V$, is given by the matrix element $A_{i,j}$.
 For an unweighted graph, $\operatorname{Haf}(A)$ can be viewed as the sum of all the perfect matchings over the graph $G$ associated with the $2M\times2M$ adjacency matrix $A$~\cite{Gupt2019TheSampling}.
 For a weighted graph, a positive weight $w_{i}$ is associated with each vertex $v_{i}$.
 Hence, the weight of a clique $C$ is defined as $w_{C}=\sum_{i \in C} w_{i}$ and the maximum-weighted clique $C_{max}$ is the largest.
 A more general and versatile approach to rescale an adjacency matrix uses a rescaling matrix $\boldsymbol{\Omega}$ leading to a rescaled matrix $A'=\boldsymbol{\Omega} A\boldsymbol{\Omega}$~\cite{Banchi2020MolecularSampling}.
 This approach is particularly useful when including node weights. $\boldsymbol{\Omega}$ is a diagonal matrix with diagonal elements defined as follows:
\begin{equation}
    \boldsymbol{\Omega}_{ii}=c(1+\alpha w_{i})
\end{equation}
where $c$ is the rescaling parameter, $w_{i}$ is the node weight of the graph and $\alpha>0$ is a constant that can be tuned to favour heavy cliques~\cite{Banchi2020MolecularSampling}.
 Rescaling $A$ is necessary so that the Takagi-Autonne decomposition returns singular values that are compatible with the accessible level of squeezing that can be reached by a given physical machine, i.e., $0\leq\tanh{r_{i}}\leq\tanh{r_{\text{max}}}\leq 1$.
 For D-GBS, the diagonal elements of $B$ will have to be rescaled accordingly: $\tilde{\boldsymbol{\gamma}}=\big(\boldsymbol{\Omega}\oplus\boldsymbol{\Omega}\big)\boldsymbol{\gamma}$.
 The only constraint on $\boldsymbol{\gamma}$ is that it must be an increasing monotonic function with respect to the weight of the node to ensure the loop weights do not introduce any bias for cliques with lower weights.

\subsection{Complexity}
The classical simulation of Gaussian Boson Sampling is widely believed to be computationally intractable. This hardness is formally proven under a few key assumptions: the Haar randomness of the unitary, the anti-concentration property of the output distribution, and that the photon statistics fall within the collision-free regime.

A Haar-random unitary ensures that the optical network lacks any underlying structure that a classical algorithm could exploit to efficiently simulate GBS. The anti-concentration property assumes that the probability mass of the output is not concentrated in a small fraction of states, but is sufficiently spread out. 

Another key assumption often used to simplify the proofs is the collision-free regime. This occurs when the probability of measuring two or more photons in a single output mode is mathematically negligible. This regime holds when the number of optical modes $M$ scales at least quadratically with the total number of photons $N$:
\begin{equation}
    M \gg N^{2}.
\end{equation}

By combining these properties, it has been shown that if an efficient classical algorithm could simulate GBS, it could be used to tightly approximate the Hafnian of a symmetric matrix. Because computing this Hafnian is known to be a \#P-hard problem, simulating GBS efficiently would firmly establish its computational hardness.

\subsection{\label{sec:level4}Max-clique search}
Max-clique search is a problem in graph theory that arises in numerous fields such as computational biology for modelling molecular interactions~\cite{Banchi2020MolecularSampling,Yu2023ADiscovery}, topological network analysis~\cite{Yu2025TopologicalProcessor}, financial fraud detection~\cite{VanVlasselaer2015Guilt-by-Constellation:Memberships,Shi2019DetectNetwork}, and social network analysis~\cite{Khodadadi2021DiscoveringAlgorithm, Lu2018CommunityConductance,Benson2016Higher-orderNetworks}.
 A max-clique is a complete subgraph with the largest number of nodes embedded in a larger unweighted graph.
 This definition can be extended to weighted graphs by defining the weight of a clique as the sum of the weights of the nodes contained in the clique.
 The max-clique search is known to be an NP-hard problem~\cite{Karp1972ReducibilityProblems}, namely an exact algorithm would run in a superpolynomial time to find the solution in the worst case.
 While exact algorithms are not efficient for the max-clique problem, like Bron-Kerbosch algorithm or branch-and-bound methods~\cite{Marino2024AAlgorithms}, heuristic algorithms aim to efficiently provide approximate solutions that will be considered satisfactory for real-world problems at a higher speed.
 The class of heuristic algorithms that can be run on a classical machine includes greedy shrinking, local search and Monte-Carlo methods~\cite{Marino2024AAlgorithms, Banchi2020MolecularSampling,Wu2015AProblems}.
 Given the connection between hafnian and graphs, GBS has been shown to be efficient to find max-cliques~\cite{Banchi2020MolecularSampling} and dense subgraphs~\cite{Arrazola2018UsingSubgraphs,Sempere-Llagostera2022ExperimentallyDevice}.
 This finding stems from the fact that a GBS machine will output a high-density subgraph with higher probability due to the correlation between density and the number of perfect matchings~\cite{Arrazola2018UsingSubgraphs}.
 For the max-clique problem in Ref.~\cite{Banchi2020MolecularSampling}, a graph is encoded by configuring the squeezing parameters and the linear optics circuitry carefully, then one detection sample is equivalent to a subgraph.

\subsection{Efficient classical methods for GBS-based graph search}
In this subsection, we review a quantum-inspired method that will be used as a benchmark and an efficient GBS sampler for non-negative graphs.
 A class of algorithms directly inspired by quantum processes has been designed to further improve the performance of classical algorithms.
 One can mention an algorithm that can be used for the same types of graph applications as GBS is being used for, i.e., $k$-dense subgraphs, max-clique search, clique-enumeration and clustering for instance.
 Oh's sampler algorithm~\cite{Oh2024Quantum-InspiredSampling} exploits the fact that for non-negative graph problems, a classical sampler based on $M^{2}$ two-photon boson samplers of $M$ modes, can efficiently sample from a probability distribution proportional to $\operatorname{Haf}(B_{\bm{n}})$.
 Since a GBS sampler has a probability distribution proportional to the square of the hafnian instead of the hafnian, the GBS sampler will outperform Oh's sampler at sampling subgraphs with a high hafnian, such as a max-clique or a dense subgraph.
 Recent works have proven the existence of classical algorithms that can simulate GBS statistics for non-negative graphs with polynomial time for weighted~\cite{Anand2025SimulatingTime} and dense unweighted graphs~\cite{Zhang2025EfficientGraphs} by using Monte-Carlo Markov chain (MCMC) algorithms.
 For the weighted graph method, which is the most efficient and general, a Cartesian product is taken between a graph $G$ and $K_{2}$, i.e., a two-node graph with one edge.
 Then, the GBS distribution for $G$ can be obtained by sampling a perfect matching from $G \square K_{2}$ with a suitable weight function and a Jerrum-Sinclair MCMC, then projecting the perfect matching onto the original vertex set. For the weighted graph algorithm, the mixing time, i.e., the number of iterations required for the Markov chain to converge to a stationary GBS distribution, scales as $\mathcal{O}(\Bar{c}mn^{4}\log^{2}(n\Bar{c}/\epsilon))$ where $m=|E|$, $n=|V|$, $\Bar{c}=\text{max}\{1,c\}$ and $\epsilon$ is the TVD between the sampling distribution and the GBS distribution.

\section{\label{sec:level3}Results}
\subsection{\label{sec:level3a}Unweighted graphs}
\begin{figure}[t]
    \centering
    \includegraphics{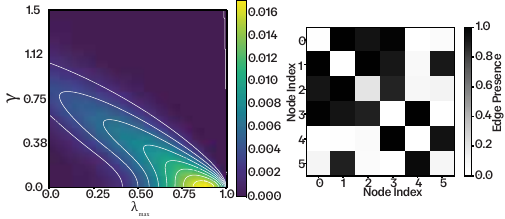}
    \caption{\textbf{Left panel}: Probability of measuring the max-clique of a graph as a function of $\gamma$ and $c$.
 \textbf{Right panel}: Adjacency matrix of the graph. }
    \label{fig:2}
\end{figure}

\begin{figure}[t]
    \centering
    \includegraphics{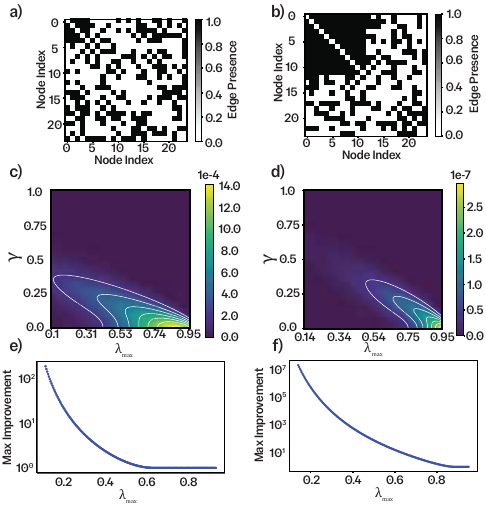}
    \caption{a) and b) Adjacency matrices.
 c) and d): Probability of measuring the max-clique of a graph as a function of $\gamma$ and $\lambda_{\text{max}}=\tanh{r_{\text{max}}}$.
 e) and f): Maximal improvement of D-GBS with respect to GBS as a function of $\lambda_{\text{max}}$ by finding the optimal $\gamma$.
 a), c) and e): 8-node clique embedded in a 20-node Erdős-Rényi graph with an edge probability of 0.2.
 b), d) and f): 14-node clique embedded in a 20-node Erdős-Rényi graph with an edge probability of 0.2.}
    \label{fig:3}
\end{figure}

One can consider a simple case with an unweighted graph.
 Since a clique is by construction a complete graph, it has the highest density, hence the highest hafnian compared to other graphs of the same size.
 Therefore, the max-clique graph is the most likely outcome over the set of subgraphs with the same size, with a probability $p(\bm{n}_{max})$.
 Adding displacement gives an additional handle with the parameter $\boldsymbol{\gamma}$, which depends on the covariance matrix and the displacement.
 The parameter $\boldsymbol{\gamma}$ represents the strength of the displacement for each mode at the output of the interferometer.
 For an unweighted graph, $\boldsymbol{\gamma}$ needs to be equal for each node such that there is no positive bias towards a specific node in the graph.
 Our metric is the probability of detecting the max-clique of known size embedded in a graph.
 We took the example of a 6-node graph with a 4-node max-clique embedded.
 This graph is depicted in the inset of Fig.~\ref{fig:2}.
 Fig.~\ref{fig:2} depicts the variation of the probability of finding the max-clique in the 6-node graph as a function of the loop strength $\gamma$ and $c$, which encodes the strength of the squeezing parameters.
 When $\gamma$ is set to zero, the setup is equivalent to a GBS configuration.
 The probability of finding the max-clique is optimal in the GBS scenario for high levels of squeezing and denoted $p_{opt}$.
 However, this high level of squeezing can be challenging to reach.
 In a practical scenario under which the optimal level of squeezing is unreachable, one can set $\gamma$ of each node to a nonzero value.
 The interpretation of this result is in the case of a low value of squeezing, the probability of producing subgraphs of the size of the max-clique is diminished, hence the postselection probability can be increased by adding displacement.
 However, the probability of finding the max-clique with $\gamma \neq 0$ will be less than $p_{opt}$ in the high squeezing regime.
 Unlike the hafnian, the loop hafnian is less optimal for finding the high density subgraphs.
 Fig.~\ref{fig:3} highlights the evolution of this wing shape as we vary the size of the clique in a graph of the same size and density.
 As the size of the max-clique increases, one also needs to increase $\gamma$ to maximise $p_{mc}$ in a low-squeezing regime.

However, for most practical applications, weighted graphs tend to be preferred against unweighted graphs since they can encode more information.
 In the next section, we investigate practical problems from Refs.~\cite{Banchi2020MolecularSampling,Yu2023ADiscovery} that require real-weighted graphs.

\subsection{\label{sec:level1b}Success rate}
In GBS, the probability of measuring a photon outcome with a fixed number of photons in a lossless case is readily given by the photon statistics of two-mode squeezed vacuum: $\ket{\tmsv}=\sqrt{1-|\lambda|^{2}}\sum_{k=0}^{\infty}\lambda^{k}\ket{k,k}$
where $\lambda=\tanh{r}\,e^{i\theta}$, $r$ is the squeezing parameter and $\theta$ is the phase of the squeezed state.
 The probability of measuring $k\geq1$ pairs of photons is $P(k)=(1-|\lambda|^{2})|\lambda|^{2k}\leq P(\text{vac})$.
 Hence, the probability of measuring vacuum is the most likely outcome of a GBS sampler.
 To increase the success rate of retrieving the max-clique given a GBS sample, we implement a routine that can be run on a classical computer, which has already been implemented in~\cite{Banchi2020MolecularSampling}.

\begin{figure}[t]
    \centering
    \includegraphics{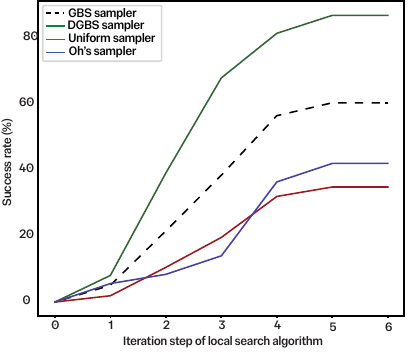}
    \caption{\textbf{Success rate of finding a max-clique} in the TaceAs~\cite{Banchi2020MolecularSampling} by using a D-GBS sampler (green) with a total displacement of $|\alpha_{tot}|^{2}=10$ and total squeezing $|\sinh{r_{tot}}|^{2}=2.34$, a GBS sampler (dashed line) with $|\sinh{r_{tot}}|^{2}=2.34$, Oh's sampler (blue)~\cite{Oh2024Quantum-InspiredSampling} and a uniform sampler (red) after running a classical post-processing routine that consists of 7 iterations of local search after greedy shrinking with 500 samples.
 }
    \label{fig:succ_rate}
\end{figure}

The classical routine is divided into two steps: greedy-shrinking and local search.
 These two stochastic methods are commonly used to find cliques in an arbitrary graph~\cite{Marino2024AAlgorithms}.
 Greedy-shrinking consists of randomly removing vertices from a given subgraph until a clique has been obtained.
 Then, $n_{iter}$ iterations of the local search routine are performed;
 each iteration consists of increasing the size of the clique by randomly adding vertices while maintaining the completeness of the subgraph and prioritizing high-weighted vertices.
 At the end of this classical routine, the success rate is computed as the number of GBS samples that led to the max-clique divided by the total number of samples.
 We simulate the GBS and D-GBS samplers with the Python libraries Strawberry Fields and The Walrus~\cite{Killoran2019StrawberryComputing,Bromley2020ApplicationsAlgorithms}.
 D-GBS and GBS are benchmarked against purely classical samplers like the uniform sampler~\cite{Banchi2020MolecularSampling} and Oh's sampler~\cite{Oh2024Quantum-InspiredSampling}.
 The first one samples subgraphs following a uniform distribution; the second one is a quantum-inspired classical sampler decomposing a GBS experiment into a set of two-photon interferences.
 This enables the algorithm to be efficient whilst mimicking a GBS experiment.
 The latter is the classical algorithm with the highest performance compared to other stochastic algorithms for finding max-cliques.
 We want to emphasize that a brute-force approach or Bron-Kerbosch are exact and deterministic algorithms at the cost of having an exponential complexity scaling.
 For instance, the worst-case complexity for an unweighted graph of size $n$ is $\mathcal{O}(3^{n/3})$ for the Bron-Kerbosch algorithm~\cite{Tomita2006TheExperiments}.
 Figure~\ref{fig:succ_rate} shows the success rate after $7$ iterations of the local search routine between a D-GBS sampler, GBS sampler, Oh's sampler, and uniform sampler.

\subsection{\label{sec:level1c}Loss resilience}
\begin{figure}[t]
    \centering
    \includegraphics{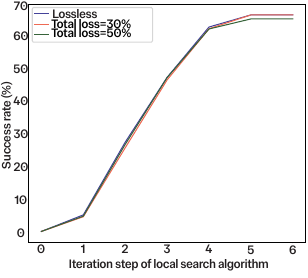}
    \caption{\textbf{Success rate of finding a max-clique in a lossy case with D-GBS} with a total displacement of $|\alpha_{tot}|^{2}=10$ and total squeezing of $|\sinh{r_{tot}}|^{2}=2.34$ in the lossless case, $\eta=0.3$ and $\eta=0.5$ with $2000$ samples.
 D-GBS is resilient to photon loss per mode with a limited decrease of $2\%$ with $50\%$ of loss per mode.}
    \label{fig:loss_mitigation}
\end{figure}

Achieving low photon loss remains challenging; previous works that claimed quantum advantage with GBS reported total loss of $69\%$ for a time-bin architecture~\cite{Madsen2022QuantumProcessor} with a single pass through the delay lines, and a $31\%$ average loss for a free space interferometer~\cite{Zhong2020QuantumPhotons}.
 High photon loss significantly degrades GBS performance by rendering it classically simulable.
 Hence, designing protocols that are resilient to loss is crucial for near-term applications in the NISQ era.
 We show in this subsection that adding photon loss in the simulation marginally diminishes the performance of the max-clique search.
 Photon loss is simulated by adding beam splitters at the output of the interferometer on each output of the unitary before the detection.
 The reflectivity of the beam splitter $\eta$ quantifies the amount of loss that is being introduced.
 Figure~\ref{fig:loss_mitigation} shows there is a marginal drop in the success rate of finding a max-clique when adding photon loss at the output of the interferometer.
 This result can be interpreted for a given clique that if some photons of the clique get lost, the resulting subgraph remains a clique and the max-clique can still be recovered after a few iterations of the local search algorithm.
 This property is also exhibited without adding displacement.

\begin{figure}[t]
    \centering
    \includegraphics{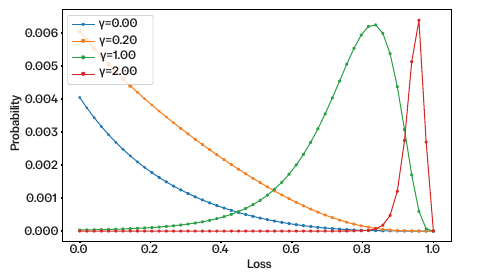}
    \caption{\textbf{Probability of finding the max-clique as a function of loss for different loop strengths $\gamma$}.
 A low level of displacement like $\gamma=0.2$ can be used to enhance the probability of the max-clique by increasing the probability of sampling the max-clique.
 For an arbitrary level of loss, the displacement can be tuned to maximize the sampling probability and outperform the ideal GBS case.}
    \label{fig:loss_mitigation_2}
\end{figure}

To explore this phenomenon, one can simulate the probability of D-GBS to sample the max-clique with photon loss.
 This computation has been performed by using the Python library Strawberry Fields.
 Photon loss $\eta$ can be modeled by placing beam splitters of reflectivity $\eta$ at the output ports of the interferometer before the detectors.
 Figure~\ref{fig:loss_mitigation_2} highlights the probability of finding the max-clique can be maximized for any value of $\eta$ by cranking up the level of displacement to compensate for the photons lost.
 This result can be simply interpreted by the fact that the displacement boosts the heralding efficiency of higher photon numbers, similar to the previous section where the amount of squeezing was insufficient to maximize the sampling probability of the max-clique.
 However, within all the subgraphs of the same size, the addition of coherent states will tend to make the probability distribution flatter, i.e., increase the entropy of the probability distribution.

\subsection{Potential Quantum-inspired algorithms}

An efficient algorithm can be designed by computing the loop hafnian efficiently for non-negative graphs by adopting an approach based on a Taylor approximation of matching-polynomials and using a chain-rule sampler to sample from this distribution~\cite{Li2025ASampling}.
 The Taylor approximation of the loop hafnian scales up as $N^{\mathcal{O}(\ln N-\ln \epsilon)}$~\cite{Li2025ASampling} and the chain-rule sampler would scale up as $\mathcal{O}(MN^{\mathcal{O}(\ln N-\ln \epsilon)})$. These results emphasize that GBS applications based on the sampling of non-negative graphs can only have a sub-exponential advantage over classical graph algorithms at best.

\subsection{\label{subsec:scalability}Scalability}

In this section, we explore the potential of scalability of a D-GBS sampler in the context of max-clique search for higher mean photon numbers and modes for potential real-world applications.

\begin{figure}[t]
    \centering
    \includegraphics{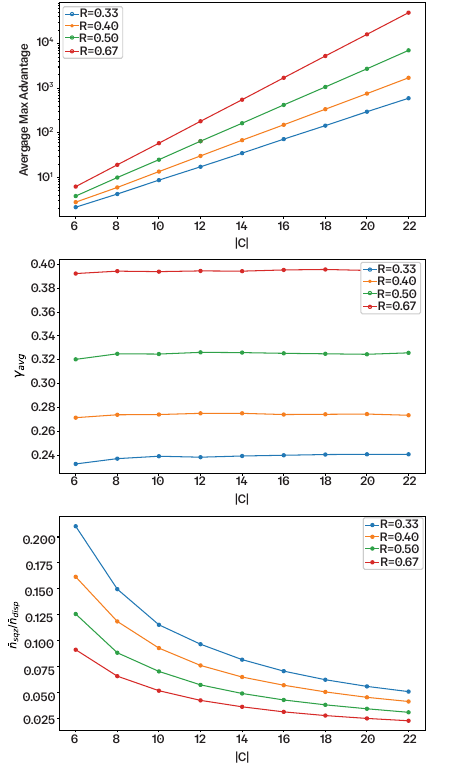}
    \caption{\textbf{Left:} Averaged enhancement of D-GBS compared to GBS with respect to the max-clique size $|C|_{max}$.
 For a given size $|C|$, the average has been computed over a dataset of 50 unweighted Erdős-Rényi graphs with an edge probability of $p=0.2$ for different graph sizes $M$.
 For each graph of the dataset, $\gamma$ has been chosen to maximize the enhancement of D-GBS with a fixed rescaling factor $c=c_{max}/2$.\;\textbf{Right:} Averaged mean photon number for the displacement, squeezing and the ratio of classical over non-classical resources.}
    \label{fig:scalability}
\end{figure}

In the asymptotic case, the mean photon number coming from displacement should scale up even slower than the mean photon number coming from squeezing.
 This requirement comes from the necessity of maintaining the relative advantage of the loop-hafnian with respect to the hafnian by keeping the normalization term low enough.
 Another motivation to keep $N_{disp}\ll N_{sqz}$ is the loop hafnian can be truncated to the first few orders of Eq.~\eqref{lhafnianexpansion}:

\begin{equation}
    \operatorname{lHaf}(\mathcal{A}_{\bm{n}})\approx\operatorname{Haf}( \mathcal{A}_{\bm{n}})+\gamma^{2}\sum_{j_{1}=1}^{2N-1}\sum_{j_{2}=j_{1}+1}^{2N}\operatorname{Haf}( \mathcal{A}_{\bm{n}-\{j_{1},j_{2}\}})
\end{equation}
with $\operatorname{Haf}( \mathcal{A}_{\bm{n}})\gg\gamma^{2}\sum_{j_{1}=1}^{2N-1}\sum_{j_{2}=j_{1}+1}^{2N}\operatorname{Haf}( \mathcal{A}_{\bm{n}-\{j_{1},j_{2}\}})$.
 While the first-leading term remains predominant, the resulting probability distribution will preserve the positive bias towards high density subgraphs whilst boosting the photon number statistics.
 This scalability discussion remains asymptotic and in an average case.
 For real-world applications, the practical advantage of D-GBS would vary based on the structure of the passive unitary and the phase and amplitude of each independent squeezer and displacer.
 Fig.~\ref{fig:scalability} shows the practical advantage obtained by graphs of size $M$ with a 14-node max-clique embedded in an Erdős-Rényi graph of edge probability $p=0.2$.
 In Fig.~\ref{fig:scalability}, one can observe that the average $\gamma$ remains at a fixed $M/|C|$ ratio over higher clique sizes.
 The advantage scales up polynomially when the size of the clique increases at a fixed ratio.
 One could imagine that a brute-force method, i.e., trying all the possible combinations of subgraphs of size $K$, would scale up $\mathcal{O}(\binom{n}{k})$.

\begin{table}
    \centering
    \begin{tabular}{|c|c|c|}
        \hline
        Sampler Algorithm &  Complexity & Accuracy  \\
        \hline
        Experimental GBS & const& N.A.\\
       
 Markov-chain GBS sampler~\cite{Anand2025SimulatingTime} &  $\mathcal{O}(\Bar{c}mn^{4}\log^{2}(n\Bar{c}/\epsilon))$& $\epsilon$  \\
        Oh's sampler~\cite{Oh2024Quantum-InspiredSampling} & $\mathcal{O}(M^{3})$& \\
        Experimental D-GBS sampler & const& N.A. \\
        Simulated D-GBS sampler~\cite{Li2025ASampling} & $\mathcal{O}(MN^{\mathcal{O}(\ln N-\ln \epsilon)})$&$\epsilon$\\
        \hline
    \end{tabular}
    \caption{Summary table of complexity between GBS and D-GBS algorithms for non-negative weighted graph applications}
    \label{tab:sampler complexity}
\end{table}

\section{\label{sec:Conclusion}Conclusion}

In this paper, we showed that displacement can be used as a classical resource 
 to boost the GBS sampling probability of cliques and dense subgraphs under real experimental conditions where photon-loss is predominant and limited levels of squeezing are accessible.
 Moreover, this classical resource maintains an edge over standard GBS when scaling up to graphs of higher size, which will be required for GBS samplers to solve real-life applications.
 While high levels of squeezing can be achieved with ppKTP waveguides~\cite{Bruno2014PulsedWavelengths,Chen2009AProcessing}, photon loss prevails in the largest GBS experiments~\cite{Madsen2022QuantumProcessor, Zhong2020QuantumPhotons}, which tends to make current large GBS experiments classically simulable~\cite{Bulmer2022TheSampling, Go2025QuantumPhotons}.
 Consequently, we do not claim an exponential speedup but rather a probable polynomial speedup.
 A potential research direction could be to quantify the average speedup of D-GBS compared to other stochastic algorithms like random sampling or Oh's sampler with the tools of quantum complexity theory.
 The main experimental challenge in our scheme is to achieve phase-locking between the different squeezers and displacers to encode the desired real adjacency matrix.
 Encoding complex-weighted graphs in a GBS setup fully exploits all the degrees of freedom allowed by interference.
 A future direction of this work could be to study if the advantage of GBS in finding a max-clique or dense subgraph can be generalized to complex graphs.
 An open question regarding quantum exponential advantage could be defined with complex-weighted graphs with GBS or D-GBS that are known to prioritize subgraphs that have a higher hafnian for non-negative graphs.
 Another motivation is this could extend the range of applications of GBS.
 Complex-weighted graphs can be employed in topological data analysis to analyze phase transitions and percolation phenomena~\cite{Yu2025TopologicalProcessor}.

\begin{acknowledgments}

We thank Changhun Oh for the helpful discussions.
 This work was supported by the Engineering and Physical Sciences Research Council (EPSRC) UK Quantum Technologies Program's hubs for Quantum Computing \& Simulation (project: EP/T001062/1) and Quantum Computing via Integrated and Interconnected Implementations (project: EP/Z53318X/1), UK Research and Innovation Future Leaders Fellowship (project: MR/W011794/1) and Guarantee Postdoctoral Fellowship (project: EP/Y029631/1), and EU Horizon 2020 Marie Sklodowska-Curie Innovation Training Network (project: 956071, `AppQInfo').
 Z.L. acknowledges partial funding support from ORCA Computing
\end{acknowledgments}

 % {\bf Contributions}
 % E.M. conceived the theory model with the help of Z.L.
 % and S.Y., E.M. coded the simulations and carried out the data analysis with the help of Z.L.
 % and S.Y, E.M. prepared the manuscript with the help of Z.L, S.Y., R.B.P. and I.A.W.
 % All authors discussed the results and reviewed the manuscript.

{\bf Competing interests}

The authors declare no competing interests.
 \vspace*{1\baselineskip}
 
\appendix

\section{\label{app:Oh's sampler}Oh's sampler}
In a boson sampling experiment, if we consider an $M$-mode boson sampling with two photons on the first mode and vacuum in the other modes, the probability $p_{i,j}$ of detecting photons in the modes $i$ and $j$ after a unitary $U$ depends on the permanent $\text{Per}(U)$:
\begin{equation}
    p_{i,j}=\left\{\begin{array}{ll}
        2|U_{1,i}|^{2}|U_{1,j}|^{2} & \mbox{if}\quad i\neq j \\
        4|U_{1,i}|^{4} & \mbox{if}\quad i=j
    \end{array}\right.
\end{equation}
If we consider $K$ different two-photon boson samplings with different passive unitaries $\{U^{(k)}\}^{K}_{k=1}$ with a probability $q_{k}$ to be selected.
 If we consider one trial, we pick one linear optical circuit by sampling from a normalized probability distribution $\{p_{k}\}_{1}^{K}$, with $\sum_{k=1}^{K}p_{k}=1$.
 The probability of detecting the outcome $\bm{n}=(1,\dots,1,0,\dots,0)$, i.e., $2N$ photons are detected on the first $2N$ modes:
\begin{equation}
    p(\bm{n})=2^{N}N!\operatorname{Haf}(A_{\bm{n}}),
\end{equation}
where $A$ is a non-negative $M\times M$ matrix defined as follows:
\begin{equation}
    A_{ij}\equiv\sum_{k=1}^{K}q_{k}|U_{1,i}^{(k)}|^{2}|U_{1,j}^{(k)}|^{2}=VQV^{T}=WW^{T}
\end{equation}
with $V_{ik}\equiv|U_{1,i}^{(k)}|^{2}$, $W_{ik}\equiv\sqrt{q_{k}}|U_{1,i}^{(k)}|^{2}$, $Q=\text{diag}(q_{1},\dots,q_{K})$.
 An explicit expression for these matrices has been derived in~\cite{Oh2024Quantum-InspiredSampling}.

\section{Greedy shrinking and local search algorithms}
The classical routine is divided into two steps: greedy-shrinking and local search.
 These two stochastic methods are commonly used to find cliques in an arbitrary graph~\cite{Marino2024AAlgorithms}.
 Greedy-shrinking consists of randomly removing vertices from a given subgraph until a clique has been obtained.
 Then, $n_{iter}$ iterations of the local search routine are performed;
 each iteration consists of increasing the size of the clique by randomly adding vertices while maintaining the completeness of the subgraph and prioritizing high-weighted vertices.
 At the end of this classical routine, the success rate is computed as the number of GBS samples that led to the max-clique divided by the total number of samples.
 The greedy shrinking algorithm consists of removing nodes from a given subgraph until a clique has been found.
 The method consists of removing vertices of smallest weights until a clique is found.
 The full procedure is described in Algorithm~\ref{alg:greedy_shrinking}. The resulting clique is fed into the local search algorithm, which consists of expanding the given clique to find a larger clique.

\begin{algorithm}
\caption{Greedy Shrinking(G,S)}
\label{alg:greedy_shrinking}
\begin{algorithmic}
\Require a graph $G=(V,E)$
\Require a GBS subgraph $S$ 
\While{$S$ is not a clique}
\State $\nu \gets V_{min}$ \Comment{$V_{min}$ is the set of vertices in S with the smallest degree}
\State $w \gets \text{min}(V_{min})$ \Comment{$w$ is a subset of $V_{min}$ with minimal weight}
\If{$|w|=1$}
    \State $S \gets S \setminus \{w\}$
   
\Else
    \State $v \gets \text{rand}(w)$ \Comment{one element of $w$ is randomly picked} 
     \State $S \gets S \setminus \{v\}$
\EndIf
\EndWhile
\end{algorithmic}
\end{algorithm}
Since the output clique of the greedy-shrinking algorithm may not be the largest clique, the local search algorithm adds other vertices to the current clique 
 by random search and checks if the larger subgraph is a higher-weighted clique or not.
 The first stage of the local search is the grow stage, which consists of finding all the vertices that are connected to all the vertices in the clique; if this set is not empty then we can add all these elements to the current clique to expand it.
 If this set is empty, then the algorithm proceeds with the stage called the swap stage.
 The idea of this stage is to generate the set of all vertices in the clique except for one node labeled $v$. The algorithm randomly selects one external node and swaps it with $v$.
 This new clique has the same size as the original clique but with a slightly different set of nodes.
 This swap test enables to explore the space of solutions to find the max-clique.
 The local search procedure is based on dynamic local search~\cite{Pullan2006DynamicProblem} and phased local search~\cite{Pullan2006PhasedProblem}, which are high performing classical algorithms for the max-clique problem~\cite{Wu2015AProblems}.
 The algorithm is summarized in Algorithm~\ref{alg:local_search}. 

\begin{algorithm}
\caption{Local search(G,C,n)}
\label{alg:local_search}
\begin{algorithmic}
\Require a graph $G=(V,E)$
\Require a clique $C$
\Require an integer $n$
\State $S\gets C$
\For{$i \gets 1, n$}
\State $V_{|S|} \gets \text{Grow}(G,S)$
\If{$V_{|S|}\neq \emptyset$}
\State $S\gets S\cup\text{rand}(V_{cand})$
\Else
\State $v\gets \text{rand}(S)$
\State $V_{S \setminus \{v\}} \gets \text{Grow}(G,S \setminus \{v\})$
\State $v'\gets \text{rand}(V_{S \setminus \{v\}})$
\State $S\gets (S \setminus \{v\})\cup v'$
\EndIf
\EndFor

\end{algorithmic}
\end{algorithm}

\section{Complexity brute-force method}
A brute-force method of finding a max-clique of size $k$ in an unweighted graph of size $n$ is to enumerate all the possible combinations of subgraphs of size $k$, i.e., $\binom{n}{k}$ possible combinations in total.
 We can distinguish three regimes in the asymptotic regime:
\begin{enumerate}
    \item $n \gg 1$ and $n \gg k$: $\binom{n}{k}\sim \frac{(ne)^{k}}{k^{k}}(2\pi k)^{-1/2}\exp\left(-\frac{k^{2}}{2n}(1+o(1))\right)$
     \item $n=2k$, $k \gg 1$: $\binom{n}{k}\sim\frac{2^{n}}{\sqrt{4\pi n}}$
    \item $n-k, k \gg 1$: $\binom{n}{k}\sim\sqrt{\frac{n}{2\pi k(n-k)}}\frac{n^{n}}{k^{k}(n-k)^{n-k}}$
   
\end{enumerate}
In the first regime whereby the max-clique dimension is negligible compared to the clique size in the asymptotic case, the complexity tends to zero with respect to $n$. Hence, the problem becomes trivial. A  straightforward consequence of this conclusion is the fact we need to consider $k$ such that $n/k$ is fixed. Intuitively, the problem would be considered 
 the hardest when $n=2k$; in this scenario, the asymptotic complexity scales exponentially. The third case should be the symmetric case of the first case, then it should also scale up polynomially with respect to $n$ and $k$.

\section{Complexity analysis}
Max-clique search and dense subgraph search are known to be NP-hard problems, which means the computational complexity scales up exponentially~\cite{Feige2001TheProblem, Karp1972ReducibilityProblems}.
 Since GBS~\cite{Kruse2019DetailedSampling} is also known to be a \#P-hard problem, we can expect a polynomial speedup at most for a GBS-sampler to find dense subgraphs or a max-clique.
 A mapping has been demonstrated between the loop-hafnian and the matching polynomials, which is also \#P-hard to calculate~\cite{Jerrum1987Two-dimensionalIntractable, OleJ.Heilmann1972TheorySystems}.
 Moreover, non-negative matching polynomials can be calculated in quasi-polynomial time~\cite{Barvinok2016CombinatoricsFunctions}.
 The main finding in~\cite{Li2025ASampling} is if $N_{\text{coh}}$ is lower than $N_{\text{sqz}}$ then D-GBS remains hard to simulate (Sec.~\ref{subsec:scalability}).
 The result of Sec.~\ref{subsec:scalability} emphasizes that the optimal ratio $N_{\text{coh}}/N_{\text{sqz}}$ asymptotically scales in $\mathcal{O}(M^{-1/4})$, with $M$ the total number of modes.
 In this regime, we can argue D-GBS remains computationally hard to simulate.
 We want to stress the point that a max-clique search algorithm that relies on a sampler: uniform sampler, D-GBS-sampler or Oh's sampler for instance, will fall in the class of heuristic algorithms~\cite{Marino2024AAlgorithms}.
 This class of algorithm aims to prioritize the speed of the algorithm at the cost of sacrificing the accuracy of the algorithm.
 Other standard max-clique search algorithms like Bron-Kerbosch or max-and-bound methods will fall under the category of exact algorithms.
 This means Bron-Kerbosch will always find the max-clique without prior assumption with an exponential complexity.

\begin{figure}
    \centering
    \includegraphics{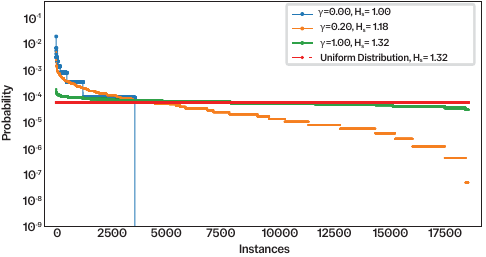}
    \caption{Histogram of probability distribution for GBS sampler, D-GBS sampler and uniform sampler based on an Erdős-Rényi graph of edge probability $p=0.2$, $c=c_{max}/2$ with 18 nodes and a max-clique of size 6. The probability distribution represents the probability of finding any subgraph of size 6 and has been normalized by the heralding probability.
 The largest probability on the left side corresponds to max-clique probability.
 As $\gamma$ increases, the probability distribution flattens, which is highlighted by the increase of the Shannon entropy.}
    \label{fig:prob_distrib}
\end{figure}

\section{Displacement for scalability}
First, we want to rewrite the normalization factor of the probability distribution for a D-GBS probability distribution to highlight its dependency with respect to $\gamma$. The argument in the exponential term of the normalization factor is Eq.~\eqref{eq:D-GBS_prob}:
\begin{align}
 \sum_{i,j}d_{i}^{*}(\boldsymbol{\Sigma}_{Q})^{-1}_{ij}d_{j}&=\sum_{i,j,\alpha\,\beta}\gamma_{\alpha}(\boldsymbol{\Sigma}_{Q})_{\alpha i}(\boldsymbol{\Sigma}_{Q})^{-1}_{ij}\gamma_{\beta}^{*}(\boldsymbol{\Sigma}_{Q})_{\beta j}^{*}\label{supp:scal1}\\
 &=\sum_{\beta j}\gamma^{*}_{\beta}(\boldsymbol{\Sigma}_{Q})_{\beta j}^{*}\gamma_{j}\label{supp:scal2}\\
 &=\hat{\boldsymbol{\gamma}}^{\dagger}(\boldsymbol{\Sigma}_{Q})^{*}\hat{\boldsymbol{\gamma}}
\end{align}
in Eq.~\eqref{supp:scal1} we make use of the fact that $d_{i}=\gamma^{*}_{j}(\boldsymbol{\Sigma}_{Q})^{*}_{ji}$ and in Eq.~\eqref{supp:scal2} the definition of matrix inversion.

\begin{equation}
    \boldsymbol{\Sigma}_{Q}^{-1}=\begin{pmatrix}
        \identity& -cA\\
        -cA&\identity
    \end{pmatrix}
\end{equation}
    
The inverted matrix can be computed by using the inversion formula for two by two block matrices stated in~\cite{Lu2002InversesMatrices}:
\begin{equation}
    \boldsymbol{\Sigma}_{Q}=\begin{pmatrix}
        \identity-cA(\identity-(cA)^{2})^{-1} & cA(\identity-(cA)^{2})^{-1}\\
        (\identity-(cA)^{2})^{-1}(cA) &(\identity-(cA)^{2})^{-1}
    \end{pmatrix}
\end{equation}
The eigenvalues of $cA$ are directly the squeezing parameters $(\tanh{r_{i}})$ from Eq.~\eqref{Takkagi}.
 Since displacement is beneficial in the regime with low squeezing or high loss, it is fair to assume that the absolute parts of the eigenvalues of $(cA)^{2}$ are all much lower than one.
 Hence, at the first order, $\identity-(cA)^{2}\approx\identity$. The Husimi covariance matrix can be simplified to:
\begin{equation}
    \boldsymbol{\Sigma}_{Q}\approx\begin{pmatrix}
        \identity-cA&cA\\
        cA&\identity\\
    \end{pmatrix}
\end{equation}
$\hat{\boldsymbol{\gamma}}=\gamma(1,1,\dots,1)^{T}$
\begin{equation}
\label{supp:scal2b}
    \frac{1}{2}\hat{\boldsymbol{\gamma}}^{\dagger}(\boldsymbol{\Sigma}_{Q}^{*})\hat{\boldsymbol{\gamma}}=\gamma^{2}\Big(M-\frac{c}{2}\sum_{ij} A_{ij}\Big)
\end{equation}
For an Erdős-Rényi graph with edge probability $p$ and $M$ nodes, each off-diagonal element of $A$ follows independent and identically distributed Bernoulli distributions.
Furthermore, the expectation value of the highest eigenvalue of an Erdős-Rényi is $Mp$ hence the right-hand side part of Eq.~\eqref{supp:scal2b} can be averaged over all Erdős-Rényi graphs $G(M,p)\in\mathcal{G}$:
\begin{align}
    \mathbb{E}_{\mathcal{G}}(\frac{1}{2}\hat{\boldsymbol{\gamma}}^{\dagger}(\boldsymbol{\Sigma}_{Q}^{*})\hat{\boldsymbol{\gamma}})&=\gamma^{2}\Big(M-\frac{M(M-1)p}{2Mp}\Big)\\
    &=\gamma^{2}\Big(\frac{M+1}{2}\Big)\\
    &=\mathcal{O}\Big(\frac{\gamma^{2}M}{2}\Big)
\end{align}

To ensure an advantage when scaling up $M$ and $N_{\text{tot}}=M\gamma^{2}+n_{\text{sqz}}$, the following condition must be met:
\begin{equation}
\label{supp:scal3}
    \gamma=\mathcal{O}\Big(\frac{n_{\text{sqz}}^{1/4}}{M^{1/2}}\Big)
\end{equation}
We can go one step further and set the constraint that the total photon number needs to satisfy the collision-free regime, i.e., $N_{\text{tot}}=M\gamma^{2}+n_{\text{sqz}}=\mathcal{O}(\sqrt{M})$. By using Eq.~\eqref{supp:scal3}, the average photon number from coherent states scales as $\gamma M^{2}=\mathcal{O}(\sqrt{n_{\text{sqz}}})$.

\bibliography{references.bib}% Produces the bibliography via BibTeX.
\end{document}